\documentclass{aastex}          
\usepackage{spr-astr-addons}    

\begin{document}
%
\title{Period-Luminosity Relations for Cepheid Variables: From Mid-Infrared to Multi-Phase}

\shorttitle{Cepheid P-L Relations}
\shortauthors{Ngeow et al.}

\author{Chow-Choong Ngeow\altaffilmark{1}} 
\author{Shashi M. Kanbur\altaffilmark{2}}
\author{Earl P. Bellinger\altaffilmark{2}}
\author{Marcella Marconi\altaffilmark{3}}
\author{Ilaria Musella\altaffilmark{3}}
\and
\author{Michele Cignoni\altaffilmark{4}}
\and
\author{Ya-Hong Lin\altaffilmark{1}}

\altaffiltext{1}{Graduate Institute of Astronomy, National Central University, Jhongli City, 32001, Taiwan}
\altaffiltext{2}{Department of Physics, State University of New York at Oswego, Oswego, NY 13126, USA}
\altaffiltext{3}{Osservatorio Astronomico di Capodimonte, Via Moiariello 16, 80131 Napoli, Italy}
\altaffiltext{4}{Department of Astronomy, Bologna University, via Ranzani 1, 40127 Bologna, Italy}

\begin{abstract}

This paper discusses two aspects of current research on the Cepheid period-luminosity (P-L) relation: the derivation of mid-infrared (MIR) P-L relations and the investigation of multi-phase P-L relations. 

The MIR P-L relations for Cepheids are important in the {\it James Webb Space Telescope} era for the distance scale issue, as the relations have potential to derive the Hubble constant within $\sim2$\% accuracy - a critical constraint in precision cosmology. Consequently, we have derived the MIR P-L relations for Cepheids in the Large and Small Magellanic Clouds, using archival data from {\it Spitzer Space Telescope}. We also compared currently empirical P-L relations for Cepheids in the Magellanic Clouds to the synthetic MIR P-L relations derived from pulsational models.

For the study of multi-phase P-L relations, we present convincing evidence that the Cepheid P-L relations in the Magellanic Clouds are highly dynamic quantities that vary significantly when considered as a function of pulsational phase. We found that there is a difference in P-L relations as a function of phase between the Cepheids in each of the Clouds; the most likely cause for this is the metallicity difference between the two galaxies. We also investigated the dispersion of the multi-phase P-L relations, and found that the minimum dispersions do not differ significantly from the mean light P-L dispersion.

\end{abstract}


\section{Introduction}\label{s:1}

The period-luminosity (P-L, also known as Leavitt Law) relation for Cepheid variables is an important astrophysical tool. A calibrated P-L relation can serve as the first rung in the extragalactic distance scale ladder, which can be used to determine the Hubble constant \citep[e.g.,][and reference therein]{fre2001,san2006,rie2011}. In the local Universe, the Cepheid P-L relation can be used to measure the distances to nearby galaxies and investigate the characteristics of our own Galaxy \citep[e.g.,][and reference therein]{maj2009,ped2009}. Research on Cepheid P-L relations includes calibrating the relations \citep[e.g.,][and reference therein]{fou2007}, investigating the metallicity dependence \citep[e.g.,][and reference therein]{rom2008} or universality of the P-L relations \citep[e.g.,][and reference therein]{bon2010}, and the study of non-linearity of these relations \citep[e.g.,][and reference therein]{nge2009}. These works mainly focused on mean light in the optical and near infrared ($JHK$) bands. In this paper, we discuss two aspects of current research in P-L relations: the extension of the P-L relations to mid-infrared (Section \ref{s:2}), and the investigation of P-L relations at various phases of the pulsation -- the multi-phase P-L relations (Section \ref{s:3}).

\section{The Mid-Infrared P-L Relations}\label{s:2}

The Hubble constant is one of the most important cosmological parameters that requires being independently determined to a high degree of accuracy and precision \citep[see, for examples,][]{hu2005,oll2007,fre2010,rie2011}. A convincing example is presented in Figure 23 of \citet{mac2006}, showing the improvement of measuring cosmological parameters when the error in the Hubble constant is reduced from $\sim10$\% to $\sim5$\%. A 2\% determination of the Hubble constant is possible to achieve via mid-infrared (MIR) distance ladder \citep{fre2010}, taking a huge advantage of the fact that extinction is negligible in the MIR. The first step in constructing the MIR distance ladder is the derivation of MIR Cepheid P-L relations. 

\subsection{The Empirical Mid-Infrared P-L Relations}\label{ss:2_1}

\begin{table}
\caption{Slopes of the MIR P-L Relations} 
\label{tab:1}
\begin{tabular}{c|cc}
\tableline  
Band   & LMC & SMC \\
\tableline
$3.6\mu \mathrm{m}$ & $-3.25\pm0.01$ & $-3.23\pm0.02$ \\
$4.5\mu \mathrm{m}$ & $-3.21\pm0.01$ & $-3.18\pm0.02$ \\
$5.8\mu \mathrm{m}$ & $-3.18\pm0.02$ & $-3.23\pm0.04$ \\
$8.0\mu \mathrm{m}$ & $-3.20\pm0.04$ & $-3.25\pm0.05$ \\
\tableline 
\end{tabular}
\end{table}

The MIR P-L relations can be derived by matching archival data from the {\it Spitzer Space Telescope} to the known Cepheids in the Magellanic Clouds. This has been done in \citet{nge2008} by matching the SAGE catalogs \citep[][single Epoch data]{mei2006} to LMC Cepheids from OGLE-II \citep[$\sim600$ Cepheids,][]{uda1999}, in \citet{nge2009} by matching the updated SAGE catalogs (two Epoch data) to LMC Cepheids from OGLE-III \citep[$\sim1800$ Cepheids,][]{sos2008}, and in \citet{nge2010} by matching the SAGE-SMC catalog \citep[][single Epoch data]{gor2010} to OGLE-III SMC Cepheids from \citet[][$\sim2600$ Cepheids]{sos2010}. Details of deriving these MIR P-L relations are given in the cited papers, and will not be repeated here. The slopes of these MIR P-L relations are summarized in Table \ref{tab:1}. It is worth pointing out that the MIR P-L relations for SMC Cepheids show a break at $\log(P)=0.4$, which is also known to exist in the optical P-L relations \citep{bau1999}, suggesting that this break is due to evolutionary effects \citep{bar1998}\footnote{This is because the evolutionary effects on this P-L break, if exist, should be independent of observed band-passes.}. Independent of the {\it Spitzer} data, the $N3$ band ($\sim 3\mu \mathrm{m}$) P-L relation for LMC Cepheids was also derived based on observations with the {\it AKARI} satellite \citep{nge2010a}. In contrast to the SAGE data, the {\it AKARI} data contains the information on time of observation, which allows for the application of random-phase correction to the single epoch data \citep[for more details, see][]{nge2010a}. The slope of the $N3$ band P-L relation was found to be $-3.25\pm0.05$, in good agreement with the $3.6\mu \mathrm{m}$ P-L slopes listed in Table \ref{tab:1}.

\begin{figure}[h]
\includegraphics[width=\columnwidth]{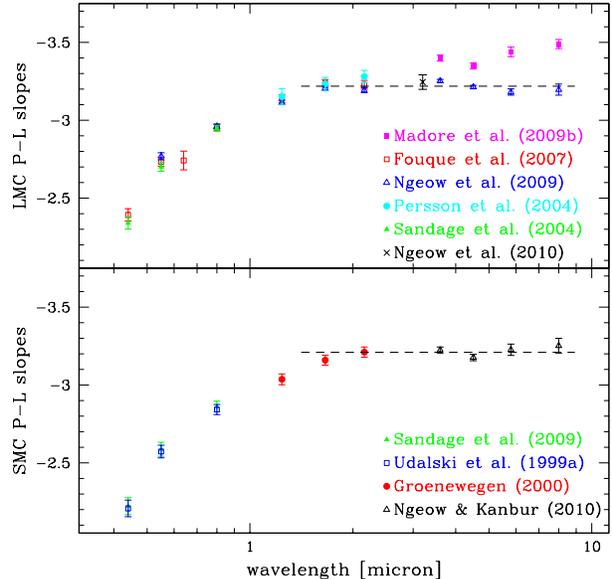}
\caption{Empirical slopes of the LMC (top panel) and SMC (bottom panel) P-L relations as a function of wavelength. The dashed lines are the expected slopes in the MIR (see text for details). [See on-line edition for a color version.]} 
\label{fig:1}
\end{figure}

\begin{figure*}[th]
\includegraphics[width=\columnwidth]{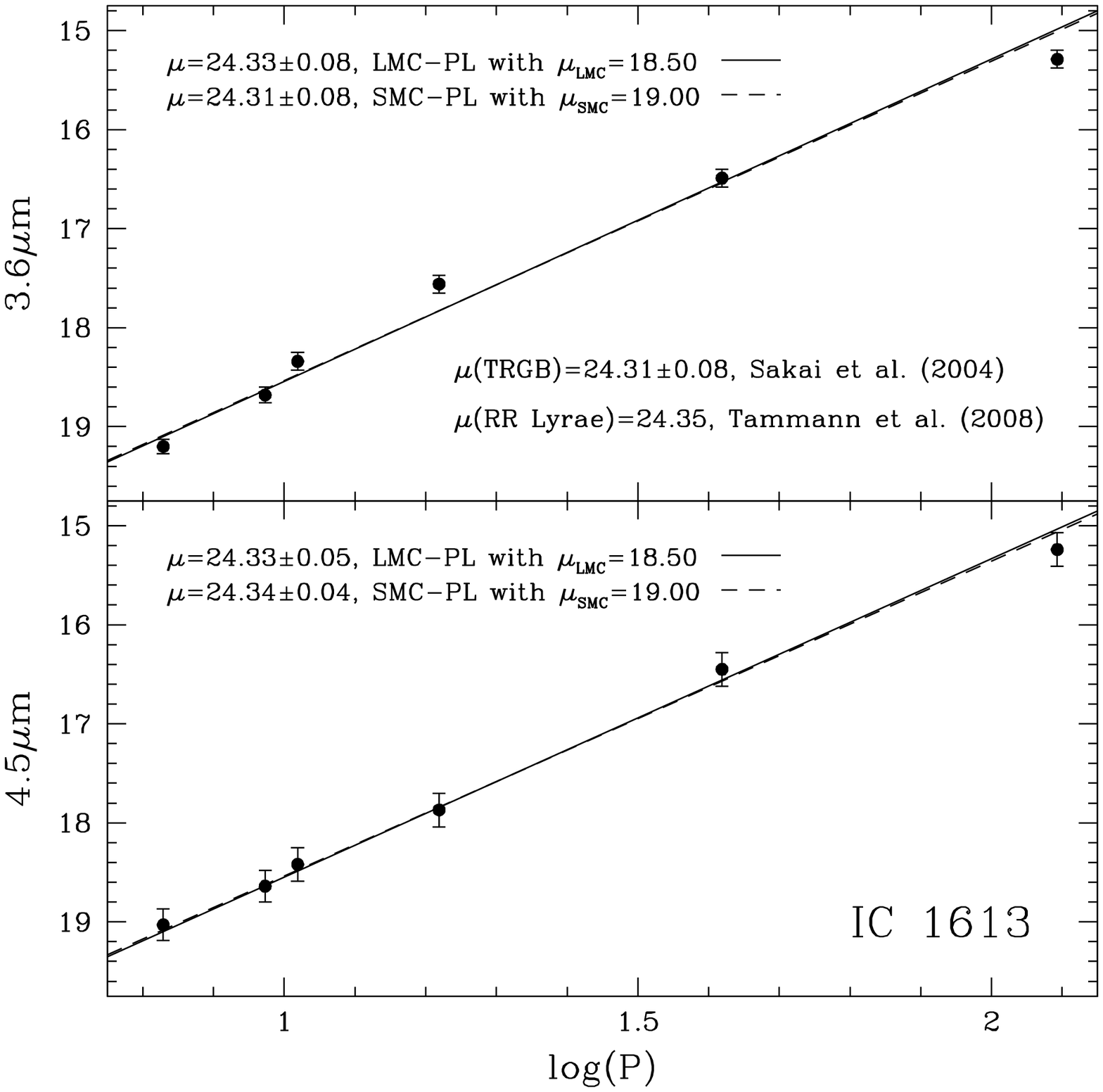}
\includegraphics[width=\columnwidth]{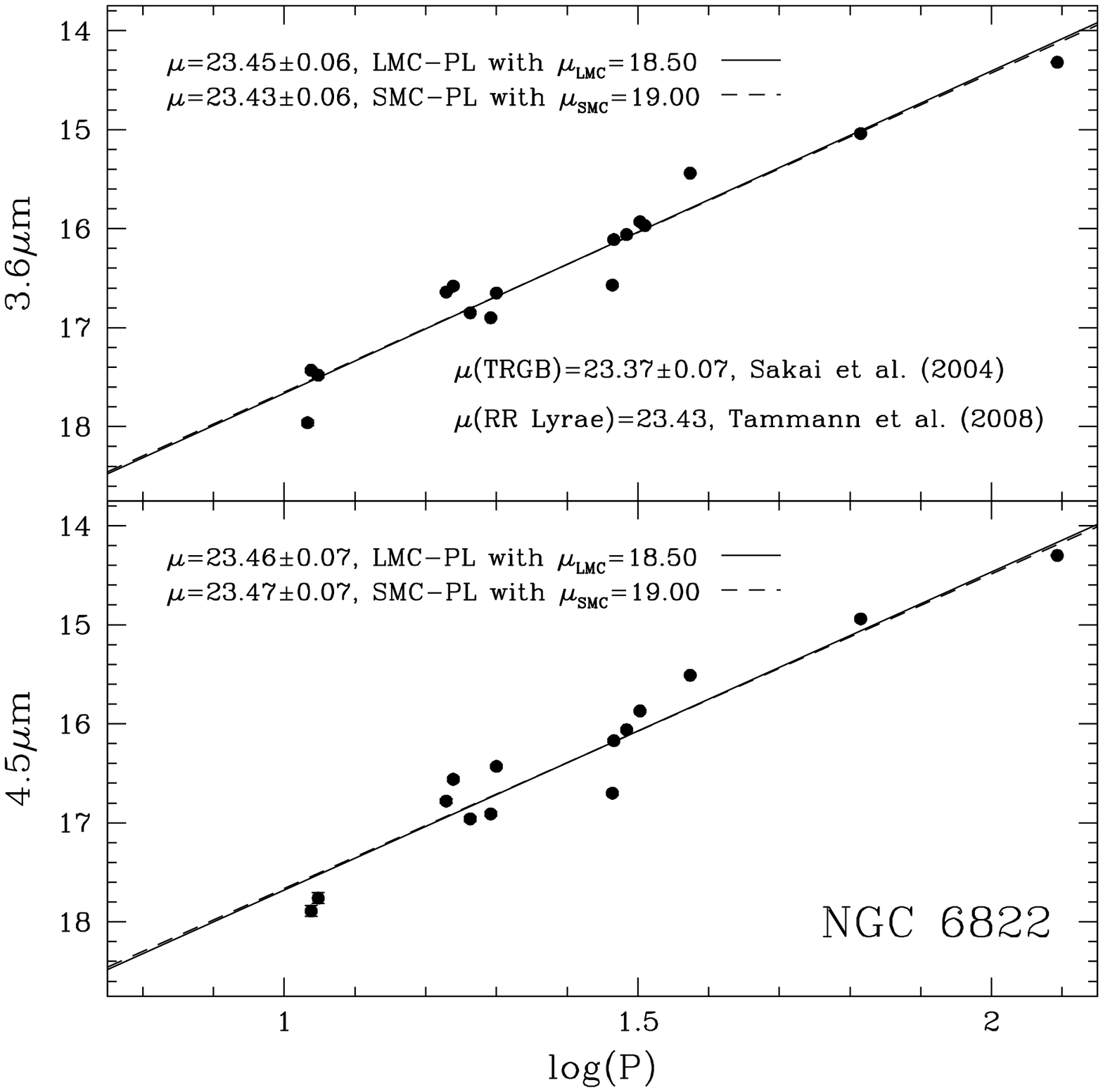}
\caption{Fitting of the empirical MIR P-L relations to the Cepheids in IC 1613 \citep[left panel, Cepheids data adopted from][]{fre2009} and NGC 6822 \citep[right panel, Cepheids data adopted from][]{mad2009a}. Distance moduli to LMC and SMC were taken to be $18.50$ and $19.00$, respectively. Note that \citet{tam2008} do not provide uncertainties in their distance moduli.} 
\label{fig:2}
\end{figure*}

Following the arguments presented in \citet{nei2010}, the slopes of the MIR P-L relations can be predicted using $L=4\pi R^2B_{\lambda}(T)$, where $B_{\lambda}(T)\propto T$ at MIR due to the Rayleigh-Jean approximation. Then, the MIR P-L relation can be written as $M_{\mathrm{IRAC}}=-5a_R\log(P)+a_T\log(P)+\mathrm{constant}$, where $a_R=0.68$ is the slope of the period-radius relation \citep{gie1999}. For $a_T$, conversion between $(V-I)$ color and temperature ($T$) was adopted from \citet{bea2001}. Using the period-color relations from Sandage et al. (2004, for LMC; 2009, for SMC), the expected slopes for the MIR P-L relations are $-3.22$ and $-3.21$ for the LMC and SMC, respectively. These values are consistent with those listed in Table \ref{tab:1}. Figure \ref{fig:1} shows the empirical P-L slopes, from the $B$ band to the IRAC bands, based on the P-L relations available in literature. The expected MIR P-L slopes are represented as dashed lines in this Figure, and suggest that the P-L slopes approach these asymptotic values around $\sim1.5 \mu \mathrm{m}$. Empirical IRAC band P-L slopes from \citet{mad2009} were included for comparison.

In parallel to the derivation of MIR P-L relations based on Magellanic Cloud Cepheids, \citet{mar2010} have also derived the MIR P-L relations from {\it Spitzer} observations for Galactic Cepheids that possess independent distance measurements in literature.

\subsection{Distance Scale Applications}\label{ss:2_2}

The empirical MIR P-L relations were used to derived the distance to two galaxies, IC 1613 and NGC 6822. The MIR photometry for Cepheids in these two galaxies were adopted from \citet{fre2009} and \citet{mad2009a}, respectively. The fitted P-L relations and the resulted distance moduli using either the LMC or SMC P-L relations were presented in Figure \ref{fig:2} for each of the $3.6\mu \mathrm{m}$ and $4.5\mu \mathrm{m}$ bands. The derived distance moduli were in good agreement when using either the LMC or SMC P-L relations, as well as between the two bands. These distance moduli were also compared to the published distance based on the Tip of the Red Giant Branch (TRGB) method from \citet{sak2004} and RR Lyrae from \citet{tam2008}. Good agreements can be found among these distance moduli as shown in Figure \ref{fig:2}.

\begin{figure}[th]
\includegraphics[width=\columnwidth]{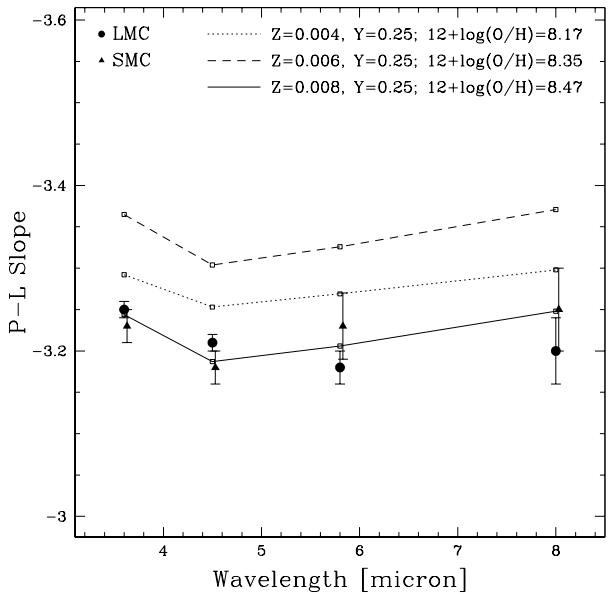}
\caption{Comparison of the empirical P-L slopes from Table \ref{tab:1} to selected synthetic P-L slopes with varying $Z$ (but at constant $Y$). Note that for better visualization, wavelengths for SMC P-L slopes have been shifted slightly.} 
\label{fig:3}
\end{figure}

\subsection{The Synthetic Mid-Infrared P-L Relations}\label{ss:2_3}

\begin{figure*}[!bh]
\includegraphics[width=\columnwidth]{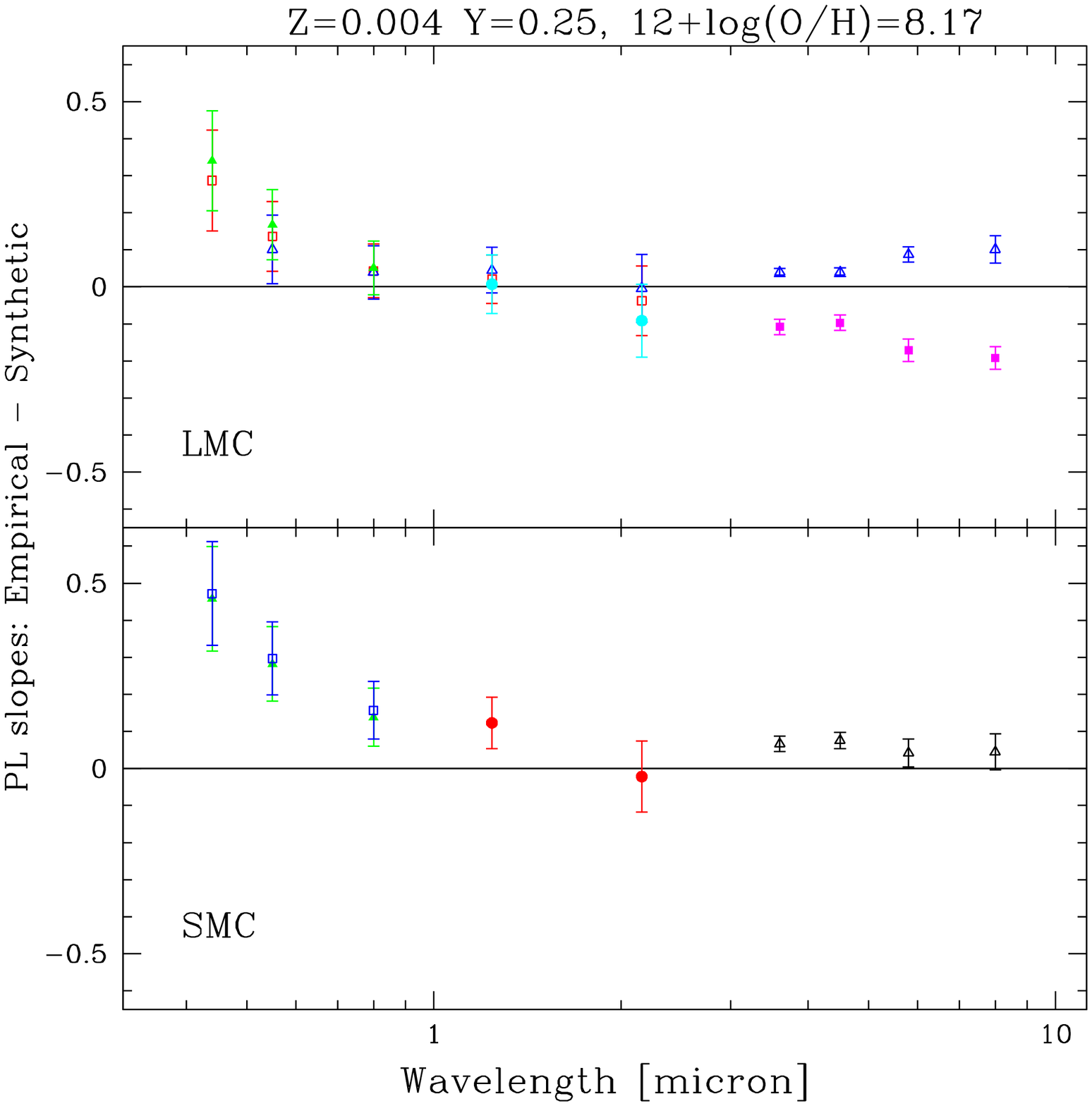}
\includegraphics[width=\columnwidth]{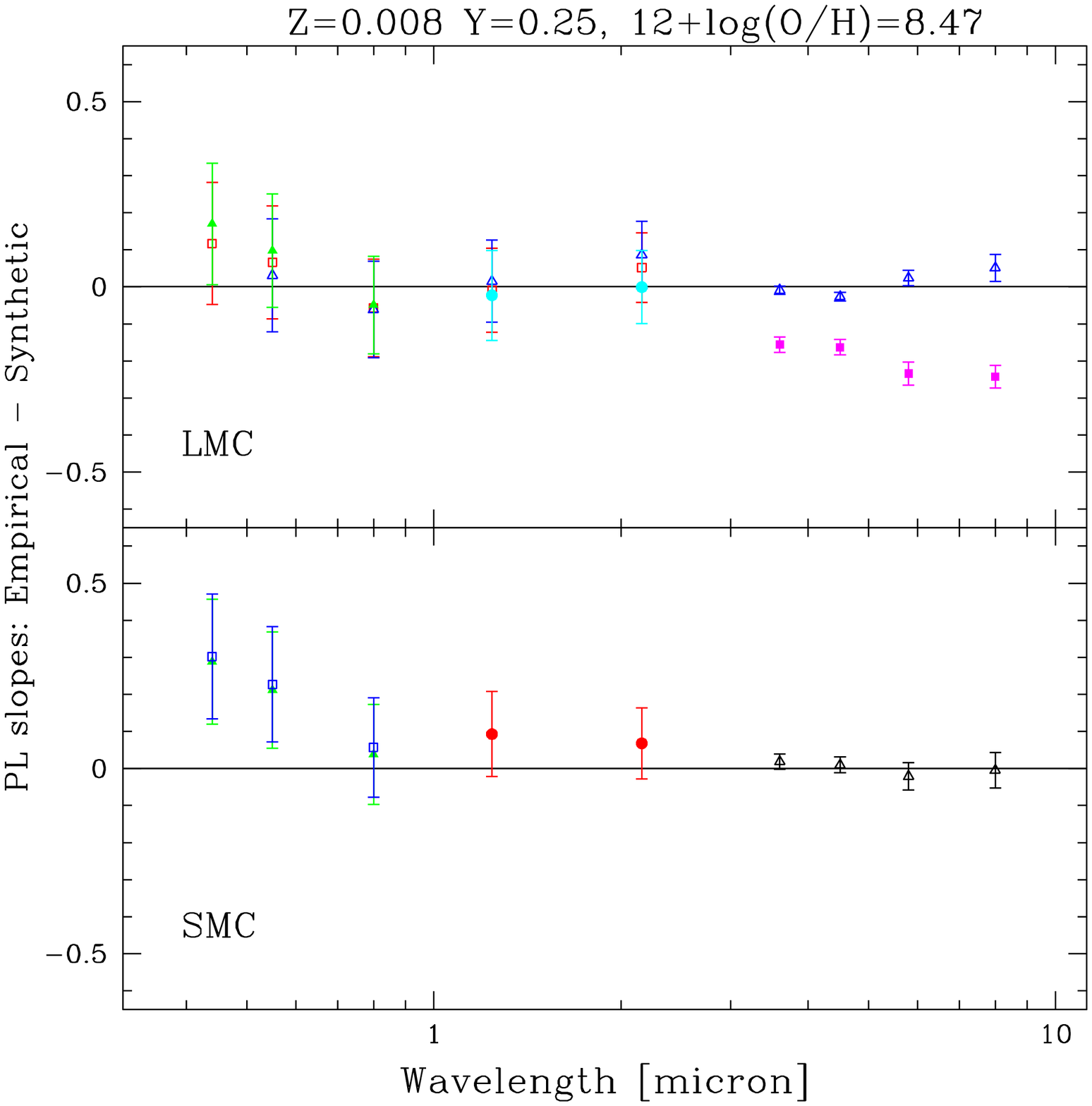}
\caption{Differences of the empirical and synthetic P-L slopes as a function of wavelength for the LMC (top panels) and SMC (bottom panels). Left and right panels are for the two model sets considered in this work. The symbols are the same as in Figure \ref{fig:1}. Synthetic P-L slopes in the $BVIJK$ bands were adopted from \citet{bon2010}. Error bars include both of the errors in empirical and synthetic P-L slopes. [See on-line edition for a color version.]} 
\label{fig:4}
\end{figure*}

A series of pulsation models with different inputs of helium ($Y$) and metal ($Z$) abundances were used to generate the synthetic P-L relations in the {\it Spitzer} IRAC bands. Details of these pulsation models and synthetic P-L relations are given elsewhere \citep{nge2011}. Briefly, non-linear pulsation codes that include time-dependent treatment of pulsation and convection, together with adopted mass-luminosity relations, were used to generate $\sim1000$ pulsators (in the mass range of $\sim5$ to $\sim11$ Solar masses) that populated the instability strip according to a given mass law \citep{ken1998}. Luminosity (and colors) of these pulsators where then converted to the IRAC band magnitudes using stellar atmosphere models. 

A comparison of the empirical P-L slopes from Table \ref{tab:1} to the synthetic P-L slopes is presented in Figure \ref{fig:3}, showing that the synthetic P-L slopes from the $(Y=0.25,\ Z=0.008)$ model set agree well to both of the empirical LMC and SMC P-L slopes. The empirical P-L slopes in various bands, as presented in Figure \ref{fig:1}, were also compared to the synthetic P-L slopes from the $(Y=0.25,\ Z=0.008)$ and $(Y=0.25,\ Z=0.004)$ model sets in Figure \ref{fig:4}, as $Z=0.008$ and $Z=0.004$ generally representing the metallicity of the LMC and SMC, respectively. For the LMC, the empirical P-L slopes are in good agreement with the synthetic P-L slopes from the $Z=0.008$ model set \citep[except for the slopes from][]{mad2009}. However, the empirical P-L slopes of the SMC agree better with synthetic P-L slopes from the $Z=0.008$ model set than the $Z=0.004$ model set. Further theoretical and empirical investigations of the SMC P-L relations are needed to solve this discrepancy.

\section{The Multi-Phase P-L Relations}\label{s:3}

Though Cepheid P-L relations are mostly studied at mean light, which is an averaged value over the pulsation cycles, P-L relations at maximum light have also been investigated in the past \citep[for example, see][]{san1968,sim1993,kan1996,kan2003}. Studies of the P-L relations beyond mean light began in a series of papers that investigated the period-color and amplitude-color relations for Cepheids at maximum, mean and minimum light \citep{kan2004,kan2004b,kan2006,kan2007}. The P-L relations at individual phases for a full pulsation cycle -- the multi-phase P-L relation -- have been studied empirically in \citet[][using OGLE-II LMC data]{nge2006}. \citet{kan2010} extended the work of \citet{nge2006} by using OGLE-III LMC data and comparisons that include predictions from theoretical pulsation models. In this Section, we continue our investigation of the multi-phase P-L relations by comparing the results found in the LMC and SMC, as well as investigating the dispersions of the multi-phase P-L relations as a function of pulsational phase.

\subsection{Data and Method}\label{ss:3_1}

Light curves data in the $VI$ bands for fundamental mode Cepheids in LMC and SMC were taken from the OGLE-III catalogs as described in \citet{sos2008,sos2010}, respectively. These catalogs also include the periods ($P$) and time of maximum light ($t_0$) of the Cepheids. Extinction corrections for the data were done by employing the extinction maps from \citet{zar2004,zar2002} for LMC and SMC Cepheids, respectively, using $R_V=3.24$ and $R_I=1.96$. The data for the $V$ and $I$ band light curves were fitted by use of a Fourier expansion in the form of:

\begin{eqnarray} 
m(\Phi) & = & m_0 + \sum_{k=1}^{n} A_k \cos[2\pi k \Phi(t) + \phi_k],
\end{eqnarray}

\noindent where $\Phi (t)=(t-t_0)/P - \mathrm{int}[(t-t_0)/P]$ is the phase of the light curves ranging from $0$ to $1$, representing a full cycle of pulsation. Hence, the P-L relation at a given phase can be derived using the magnitudes at this phase from the smooth light curves calculated using equation (1). In addition to $VI$ band multi-phase P-L relations, we also included the multi-phase relations for the extinction free Wesenheit function \citep{mad1991,uda1999a}, $W=I-1.55(V-I)$, by taking the $V$ and $I$ magnitudes at the same phase using equation (1).  
 
\subsection{Comparison of the Multi-Phase P-L Relations for Magellanic Clouds Cepheids}\label{ss:3_2}

\begin{figure*}
\includegraphics[width=3.1in]{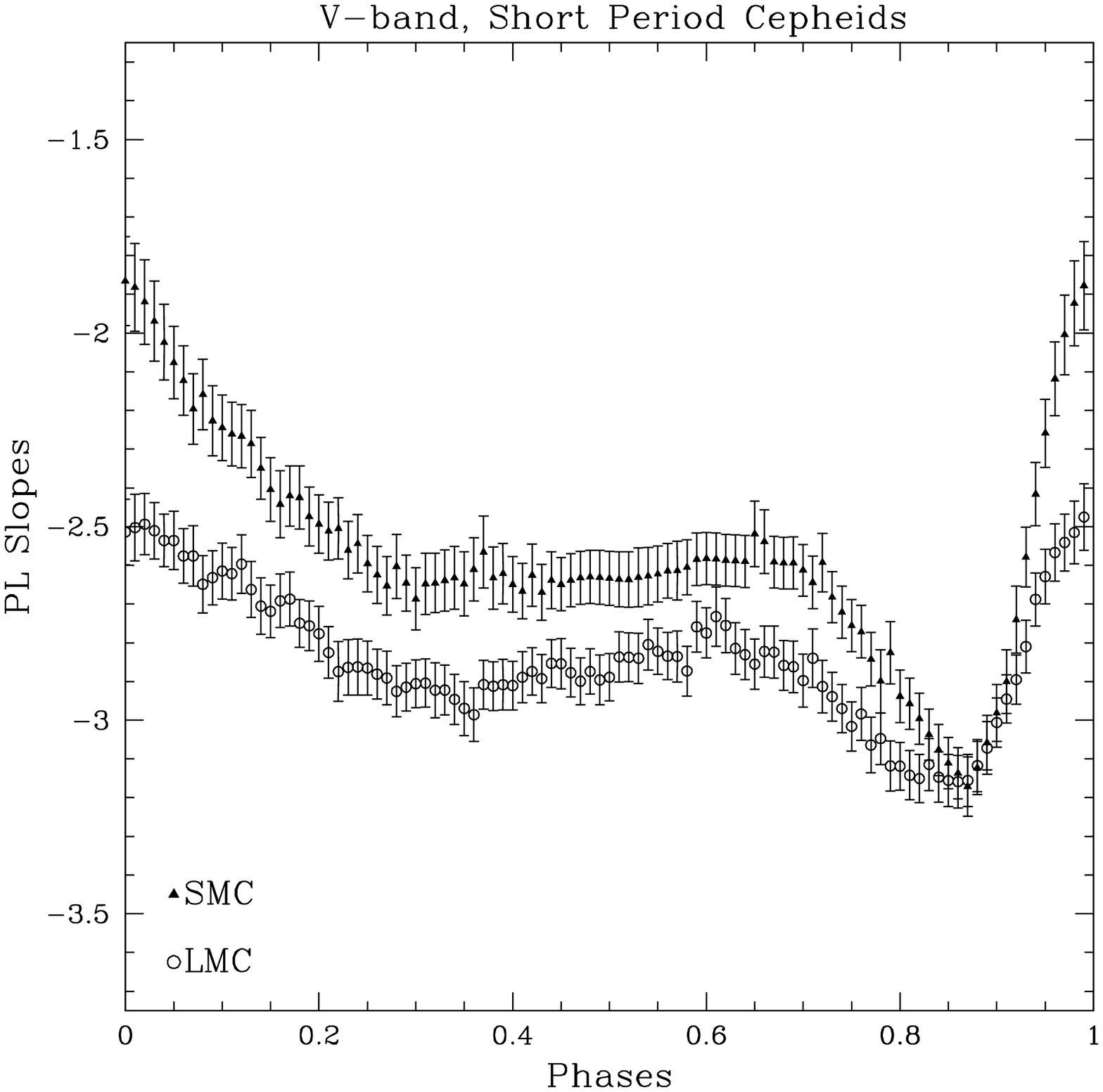}
\includegraphics[width=3.1in]{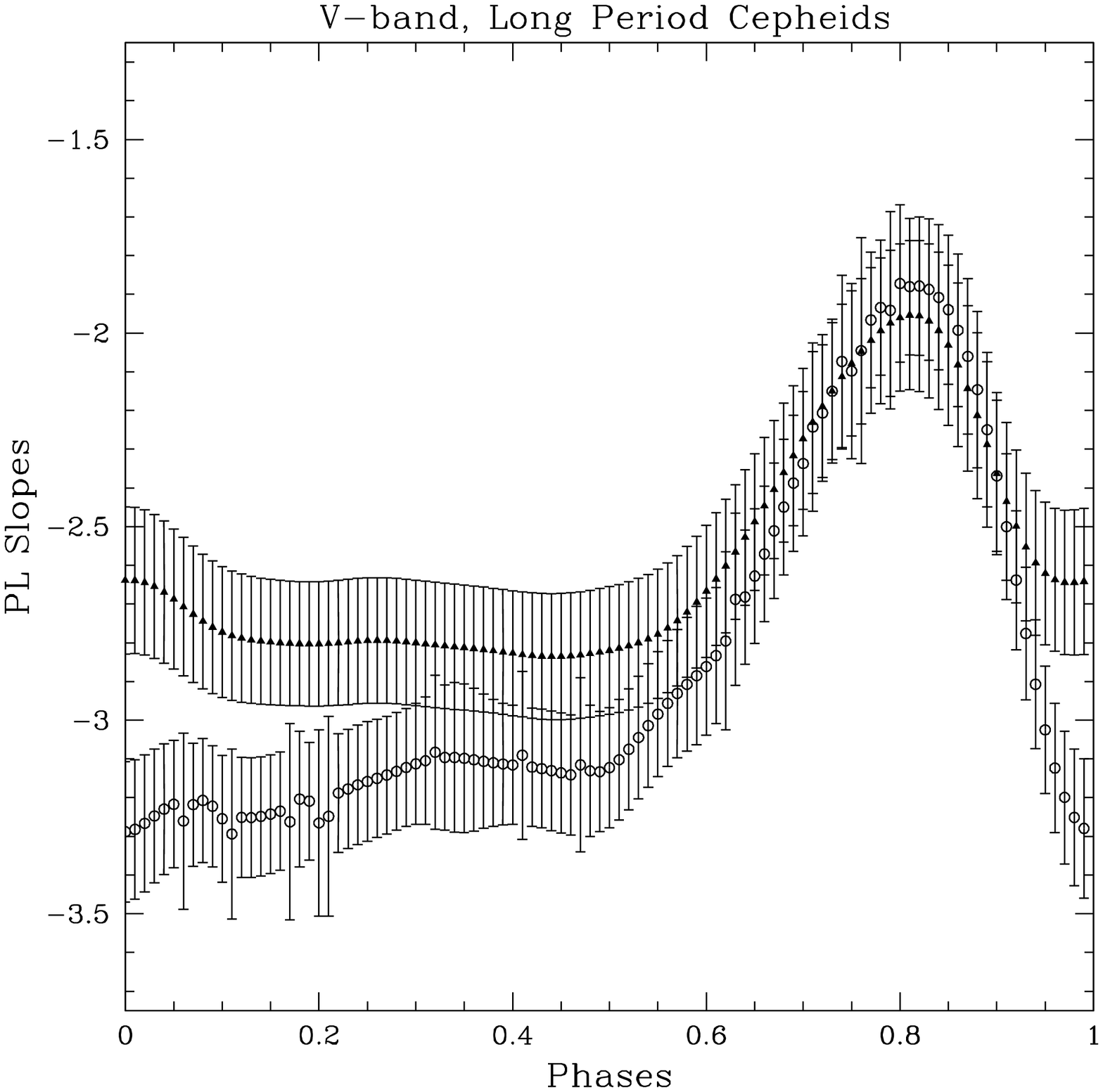} \\
\includegraphics[width=3.1in]{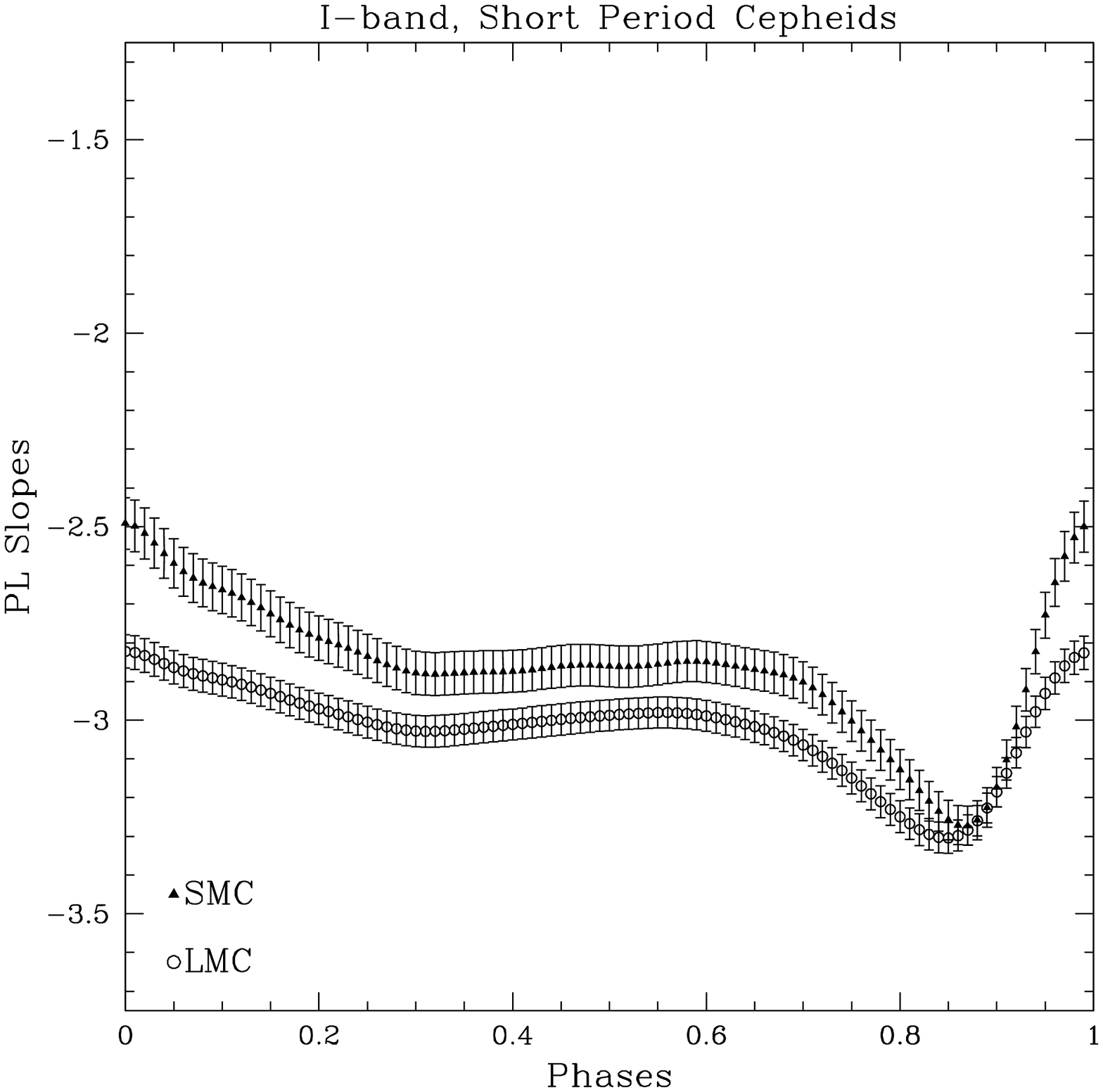}
\includegraphics[width=3.1in]{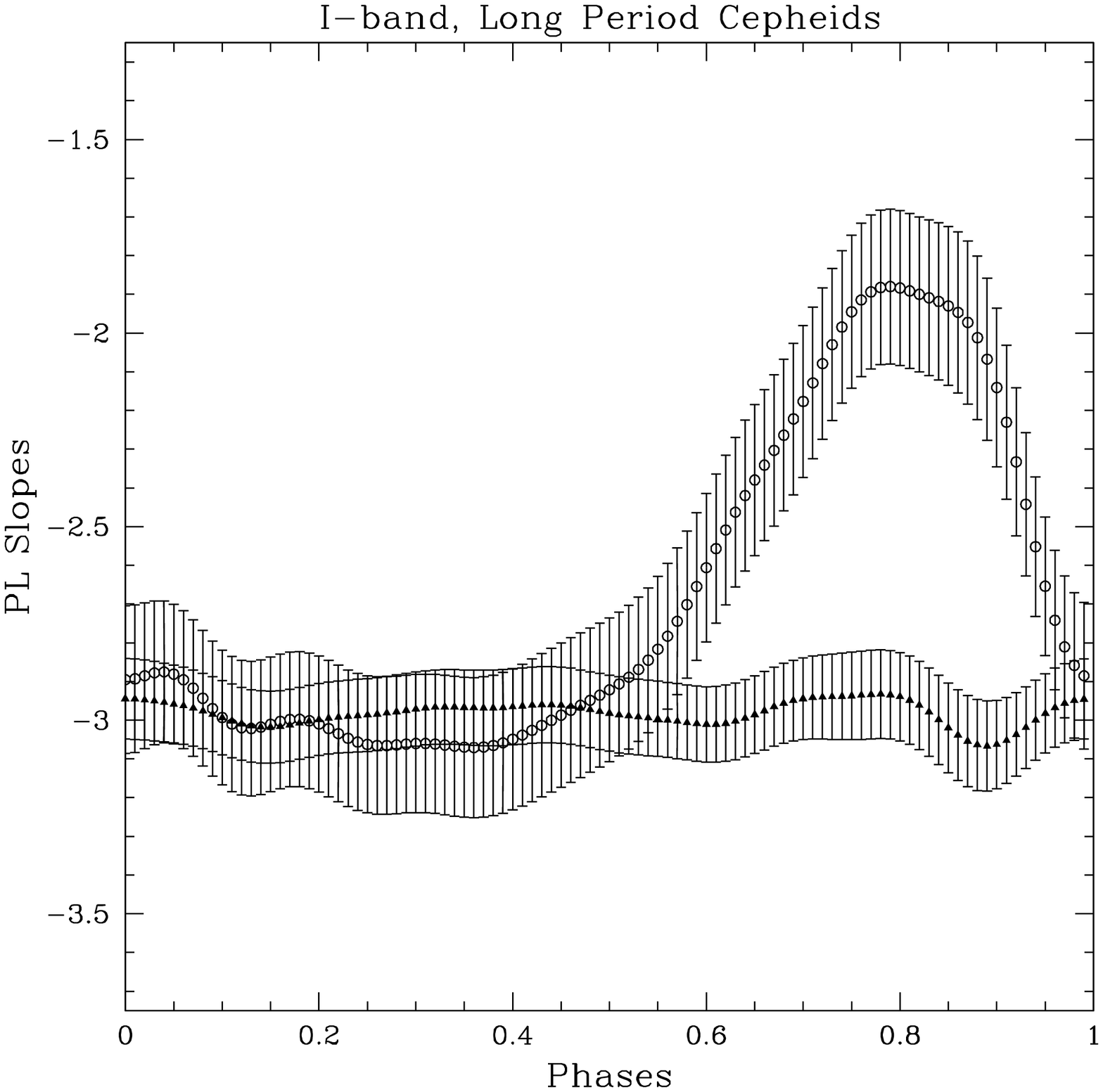} \\
\includegraphics[width=3.1in]{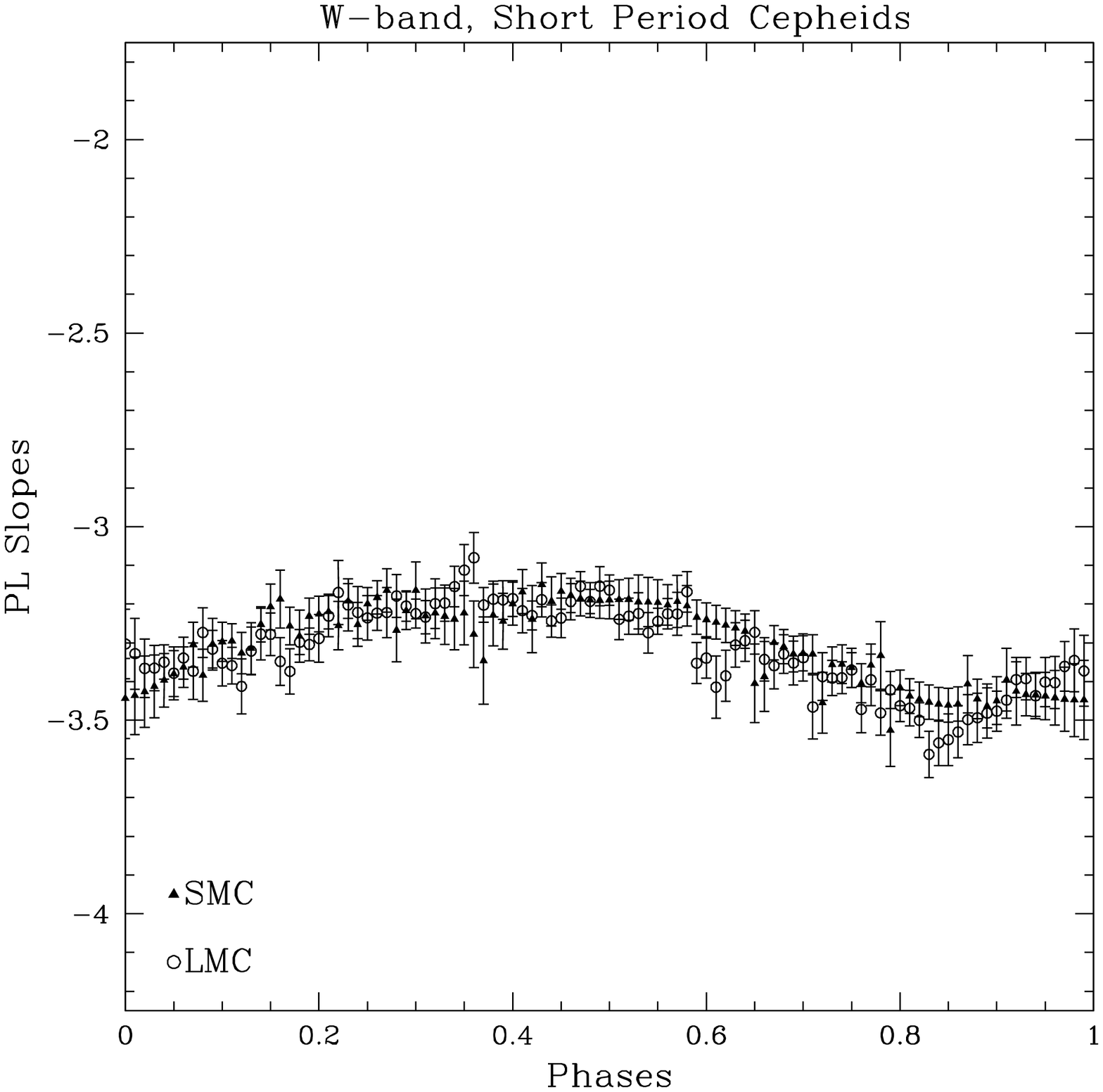}
\includegraphics[width=3.1in]{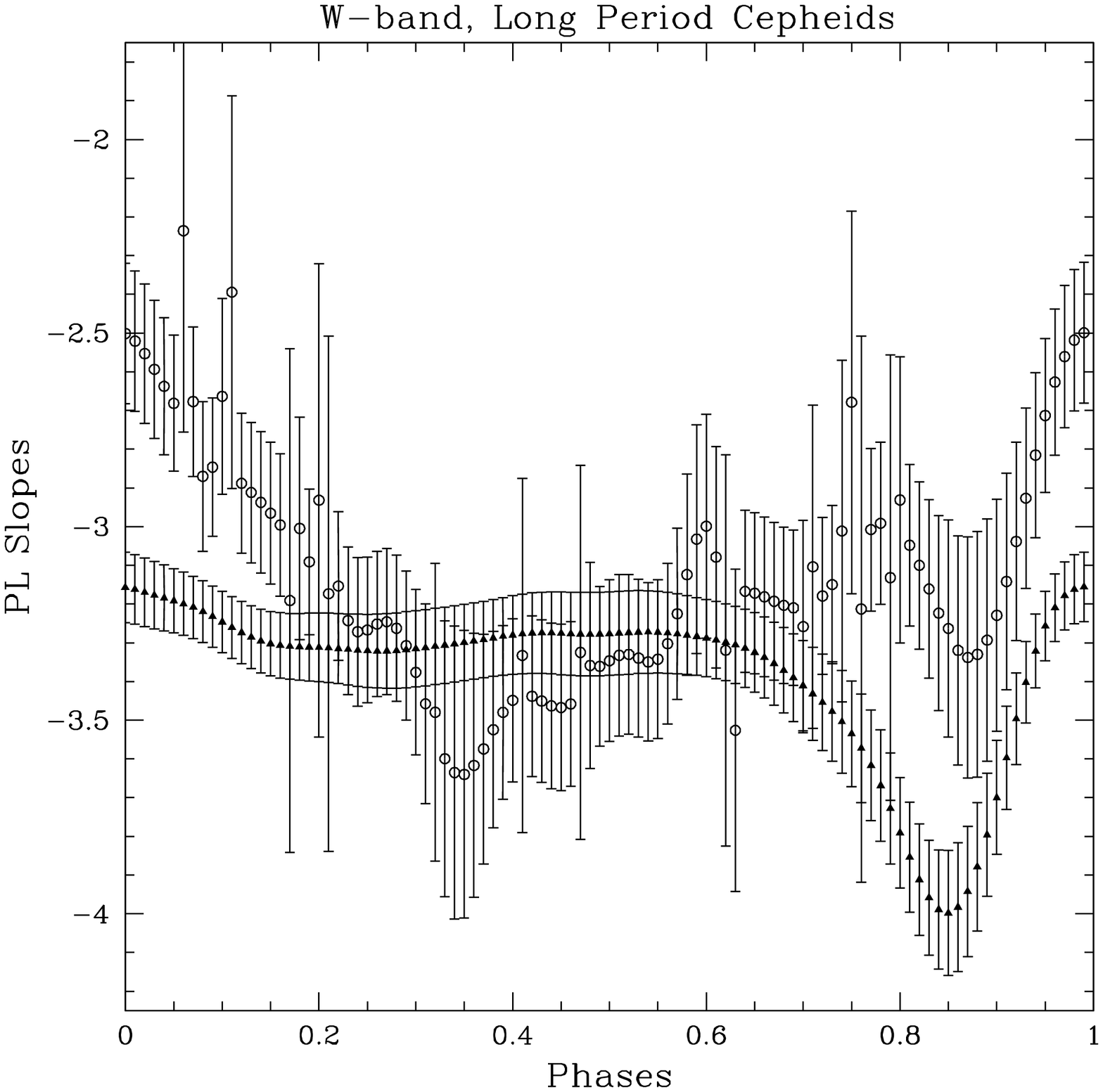}
\caption{Slopes of the P-L relations as a function of pulsational phase for LMC (open circles) and SMC (filled squares) in $V$ (top panels), $I$ (middle panels) and $W$ (bottom panels) bands. Left and right panels are for the short and long period Cepheids, respectively.} 
\label{fig:5}
\end{figure*}

\begin{figure*}
\includegraphics[width=3.1in]{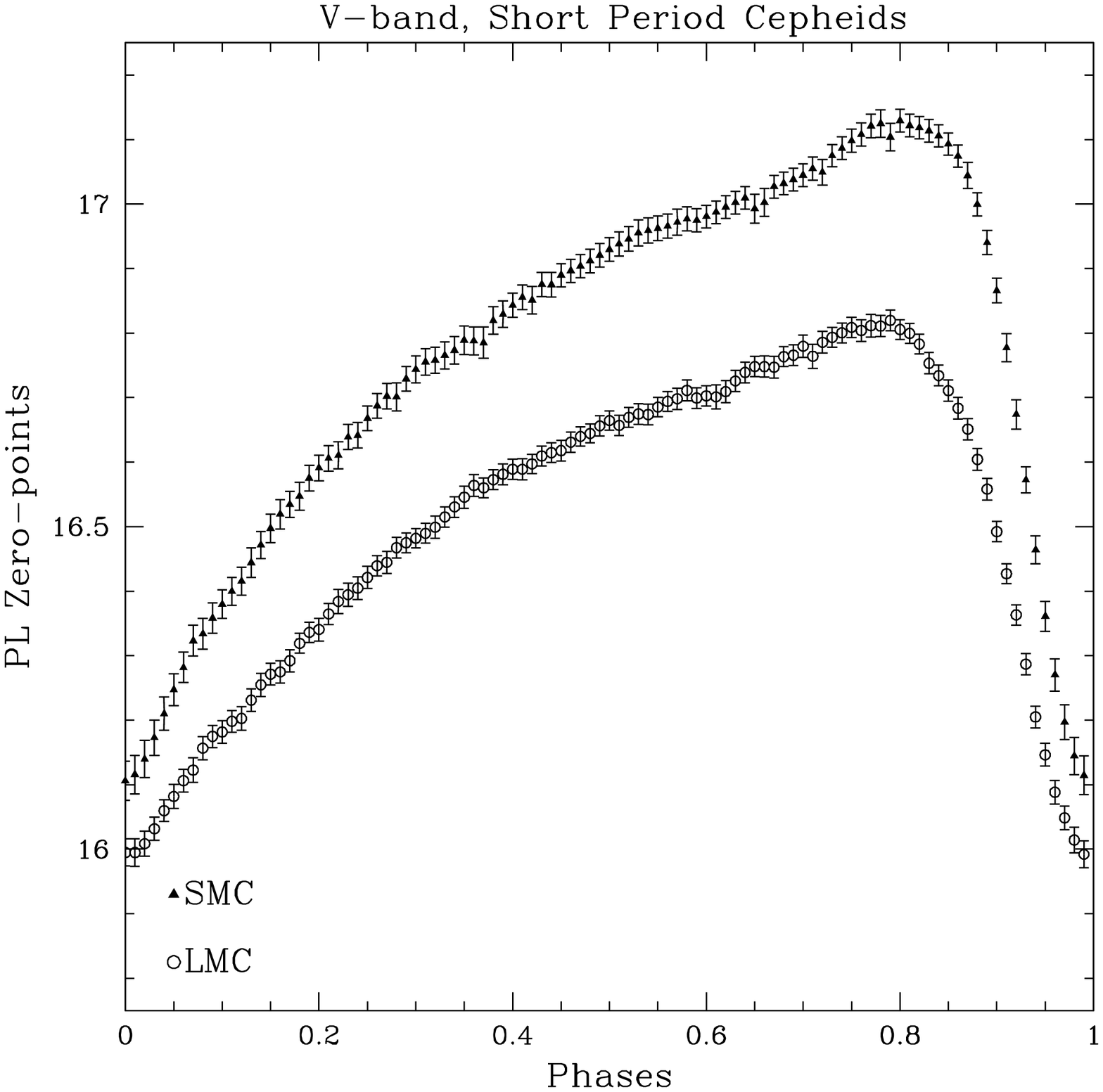}
\includegraphics[width=3.1in]{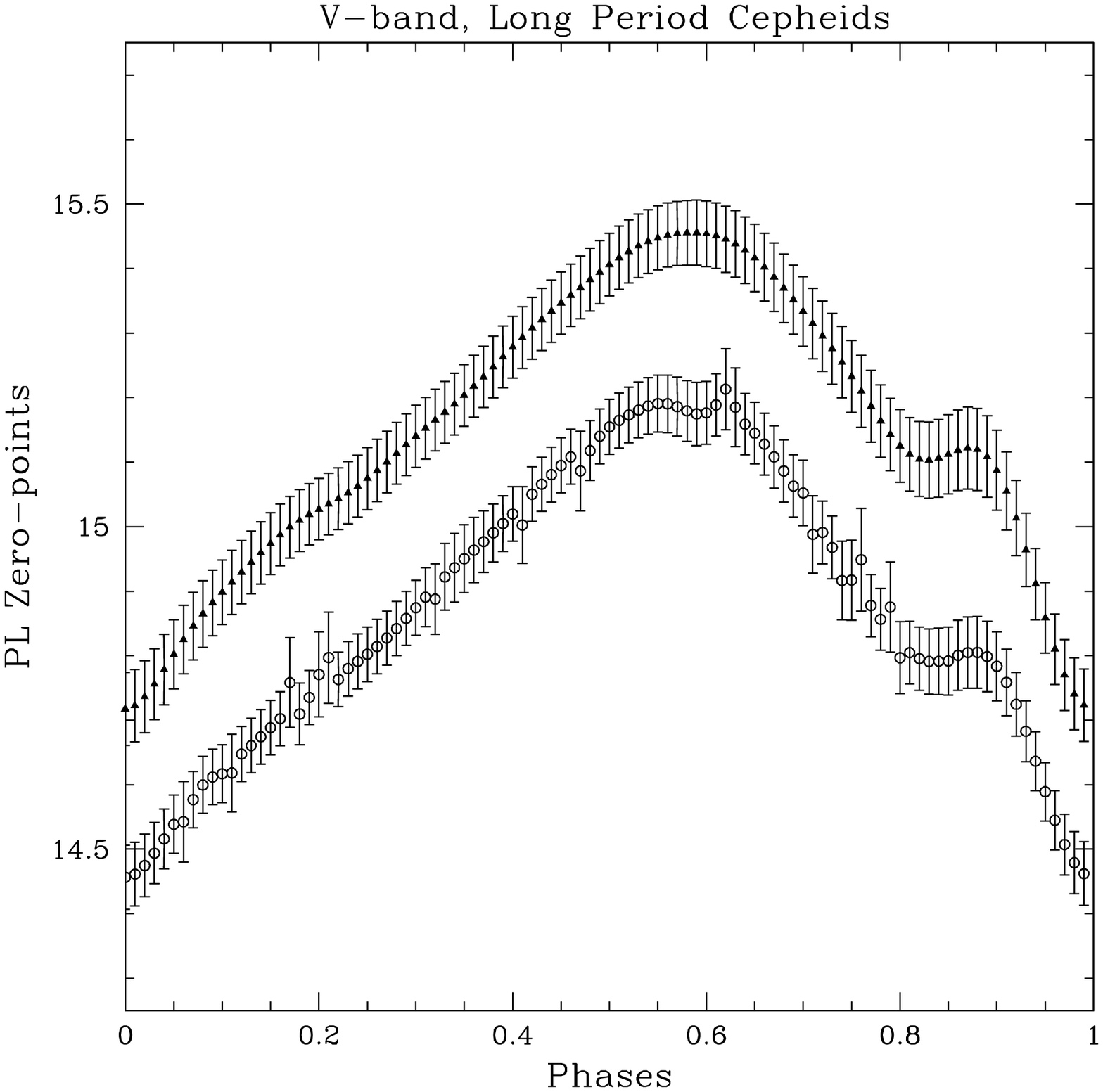} \\
\includegraphics[width=3.1in]{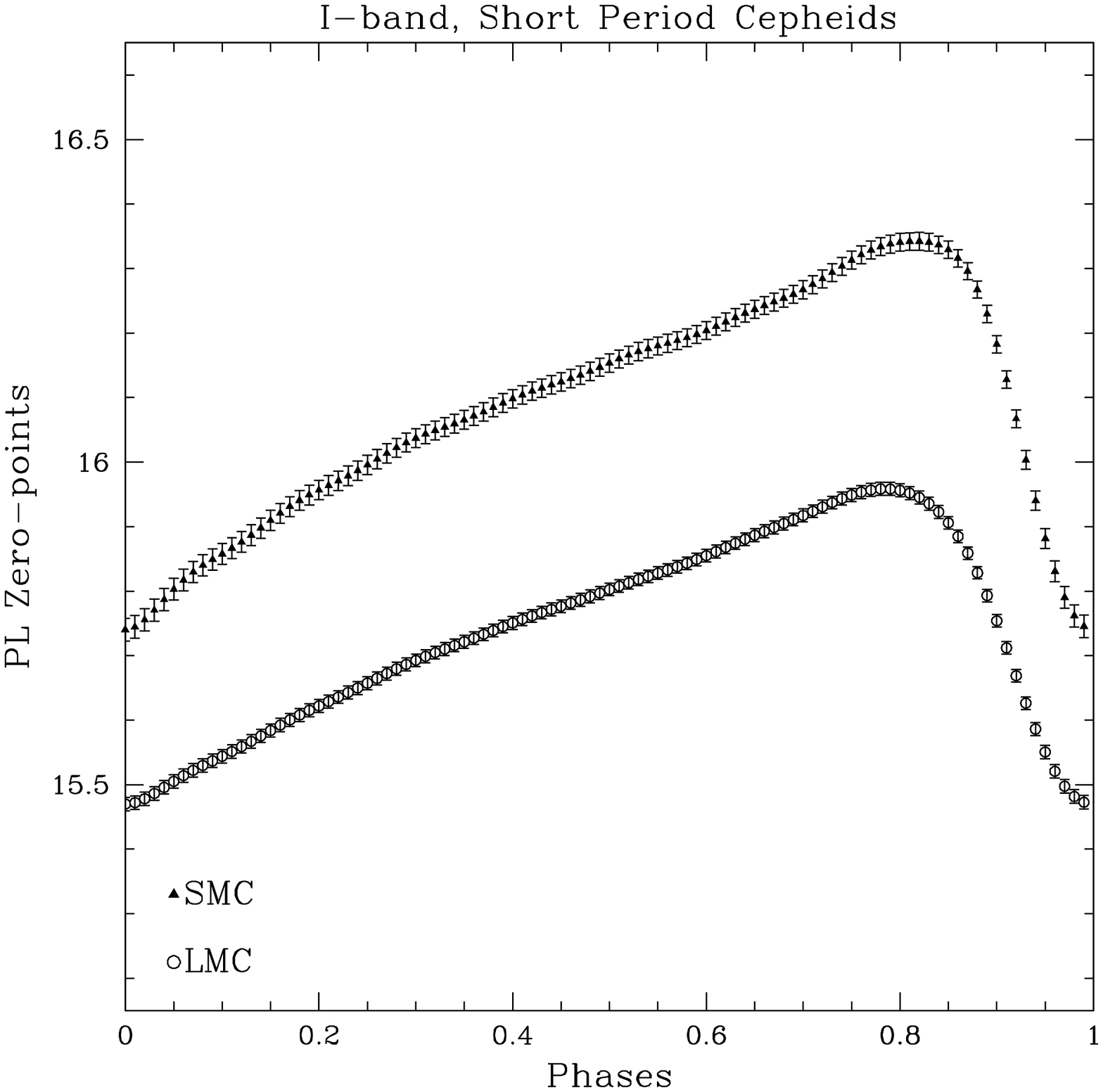}
\includegraphics[width=3.1in]{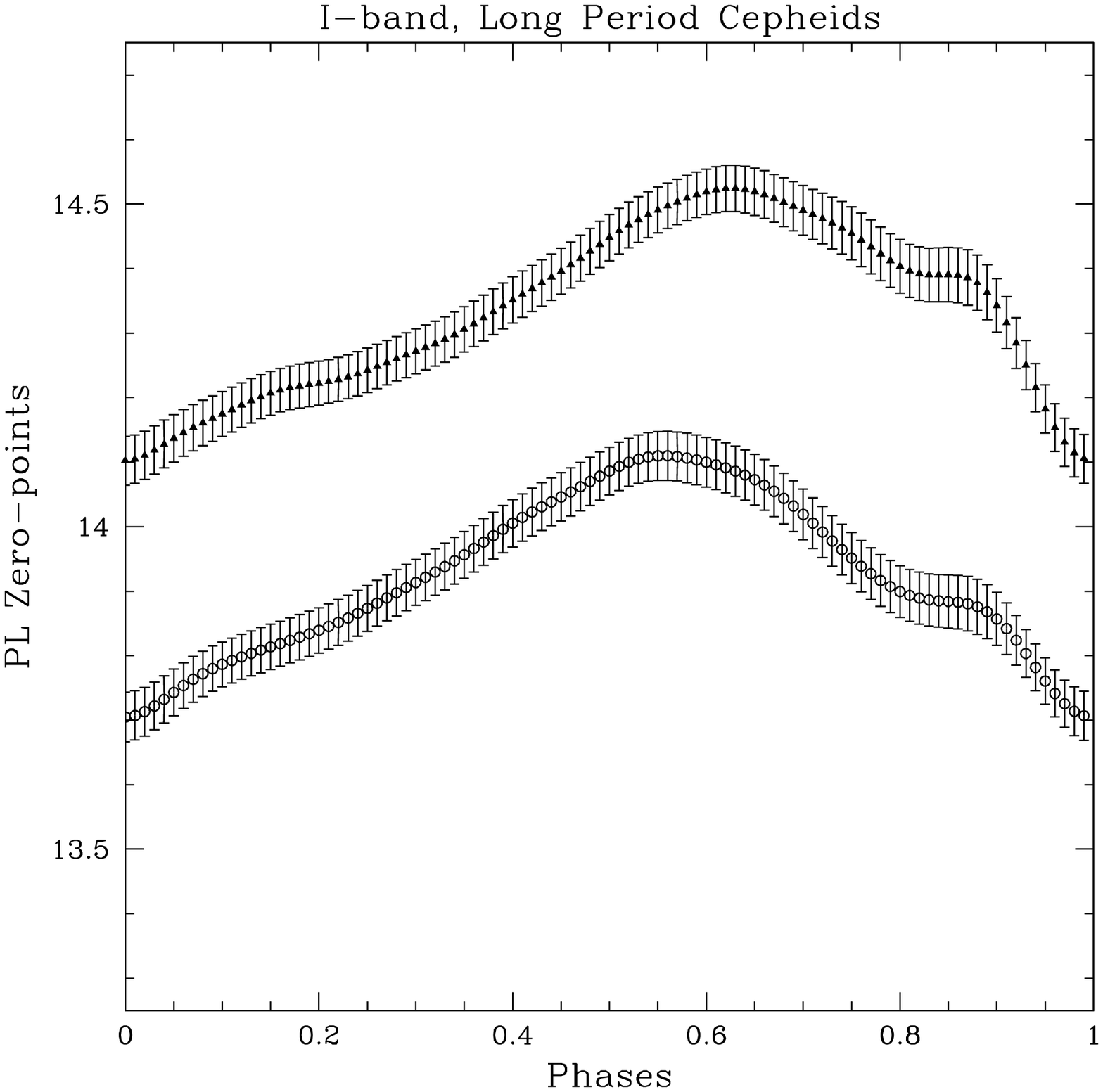} \\
\includegraphics[width=3.1in]{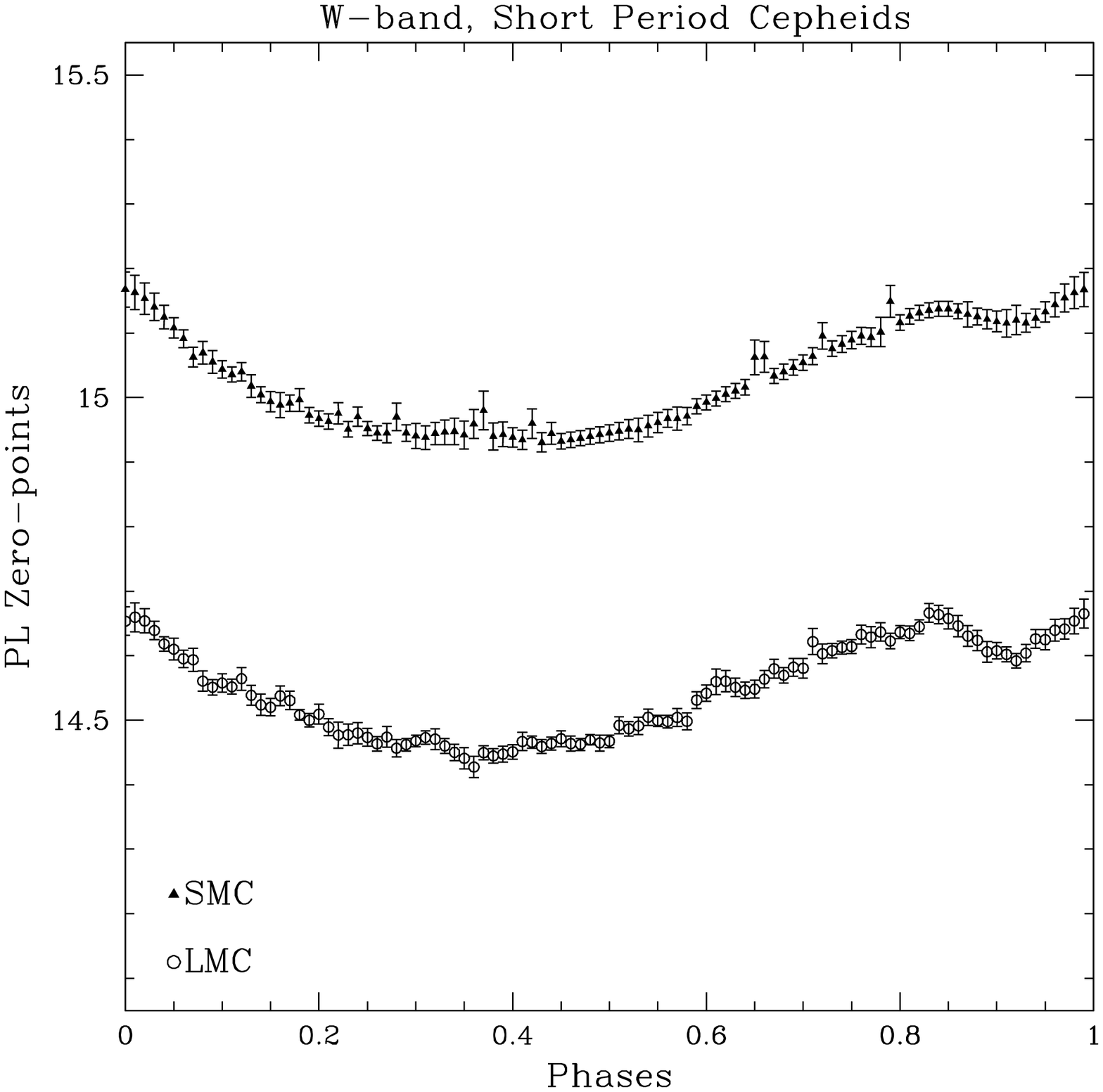}
\includegraphics[width=3.1in]{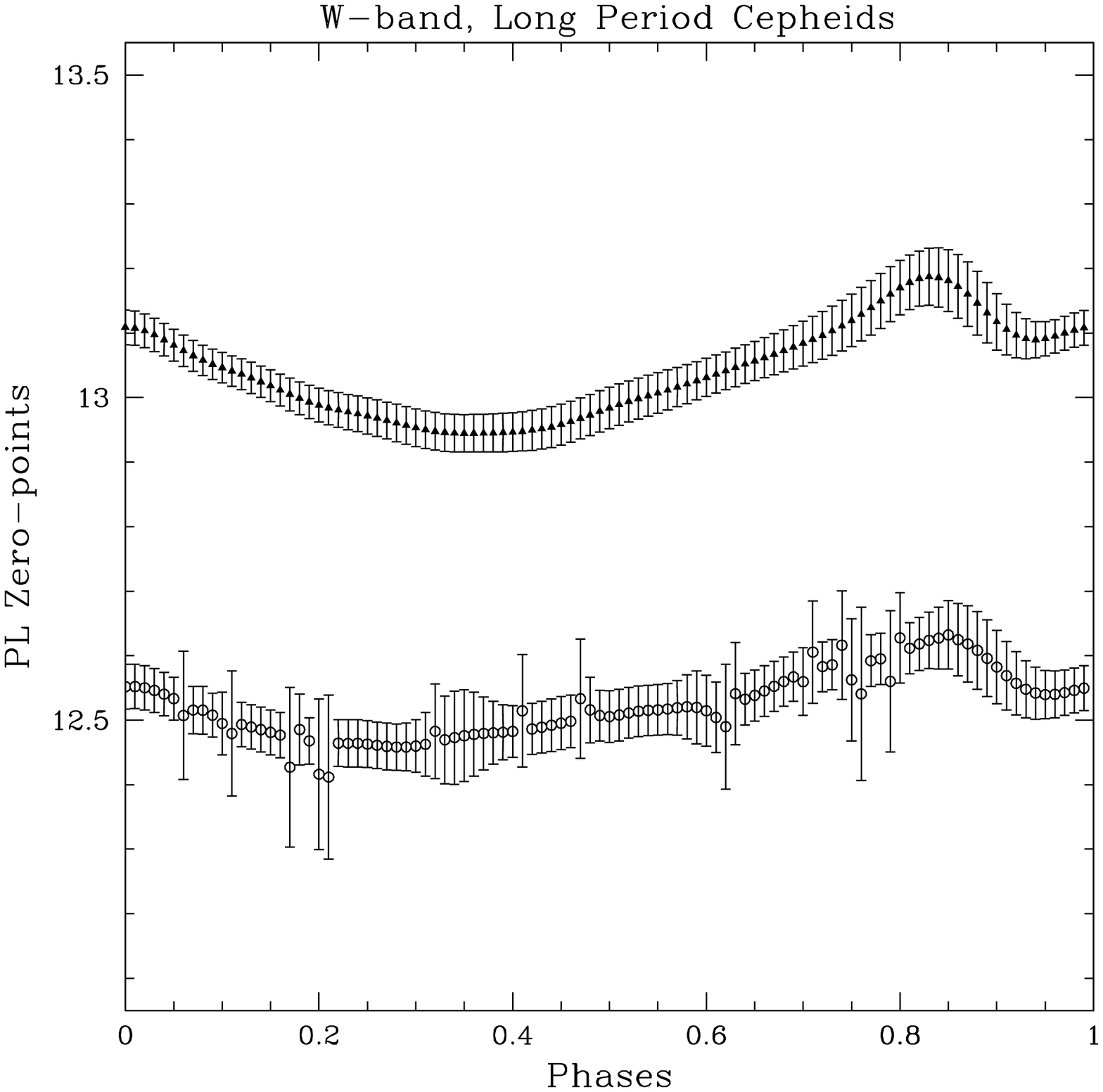}
\caption{Same as Figure \ref{fig:5}, but for the zero-points of the multi-phase P-L relations evaluated at $10$ days, i.e. fitting the P-L relations in the form of $m=a\log (P-1.0) + b$.} 
\label{fig:6}
\end{figure*}

Figures \ref{fig:5} and \ref{fig:6} present the slopes and zero-point for the multi-phase P-L relations as a function of pulsational phase for the LMC and SMC Cepheids. In these Figures, Cepheids were separated into short-period ($0.4<\log P <1.0$) and long-period ($\log P>1.0$) groups. The dynamic nature of the P-L relations as a function of pulsational phase can be seen clearly from these Figures. These results also show convincing evidence of non-linearity in the multi-phase relations, though the exact effects on the mean light P-L relations and subsequent estimates of the Hubble constant are still to be determined. Of particular interest are the differences in the long period multi-phase Wesenheit function between the LMC and SMC. This occurrence is important since the extra-galactic distance scale is primarily built with long period Cepheids. The most probable explanation for these differences is that these relations vary with metallicity. Because massive stars usually become variable after leaving the main sequence, understanding the effect that metallicity has on pulsation is important for understanding stellar evolution.

\subsection{Dispersions of the Multi-Phase P-L Relations}\label{ss:3_3}

\citet{aar1986} listed five criteria for a good distance indicator, one of them being small scatter\footnote{The other four criteria, as quoted from \citet{aar1986}, are: ``sound physical basis, quantitative observables, measurables needing minimal corrections, and applicability over a wide distance range''.}. Even though Cepheid P-L relations at mean light have been widely used in distance scale work, the dynamic nature of the multi-phase P-L relations as seen in the previous sub-section gives reason to postulate the existence of a phase at which the scatter of the respective P-L relation is smallest. If this is indeed the case, then it would be possible to improve the distance measurements, and hence the Hubble constant precision, by applying the P-L relation at this particular phase. 

\begin{figure*}
\includegraphics[width=\columnwidth]{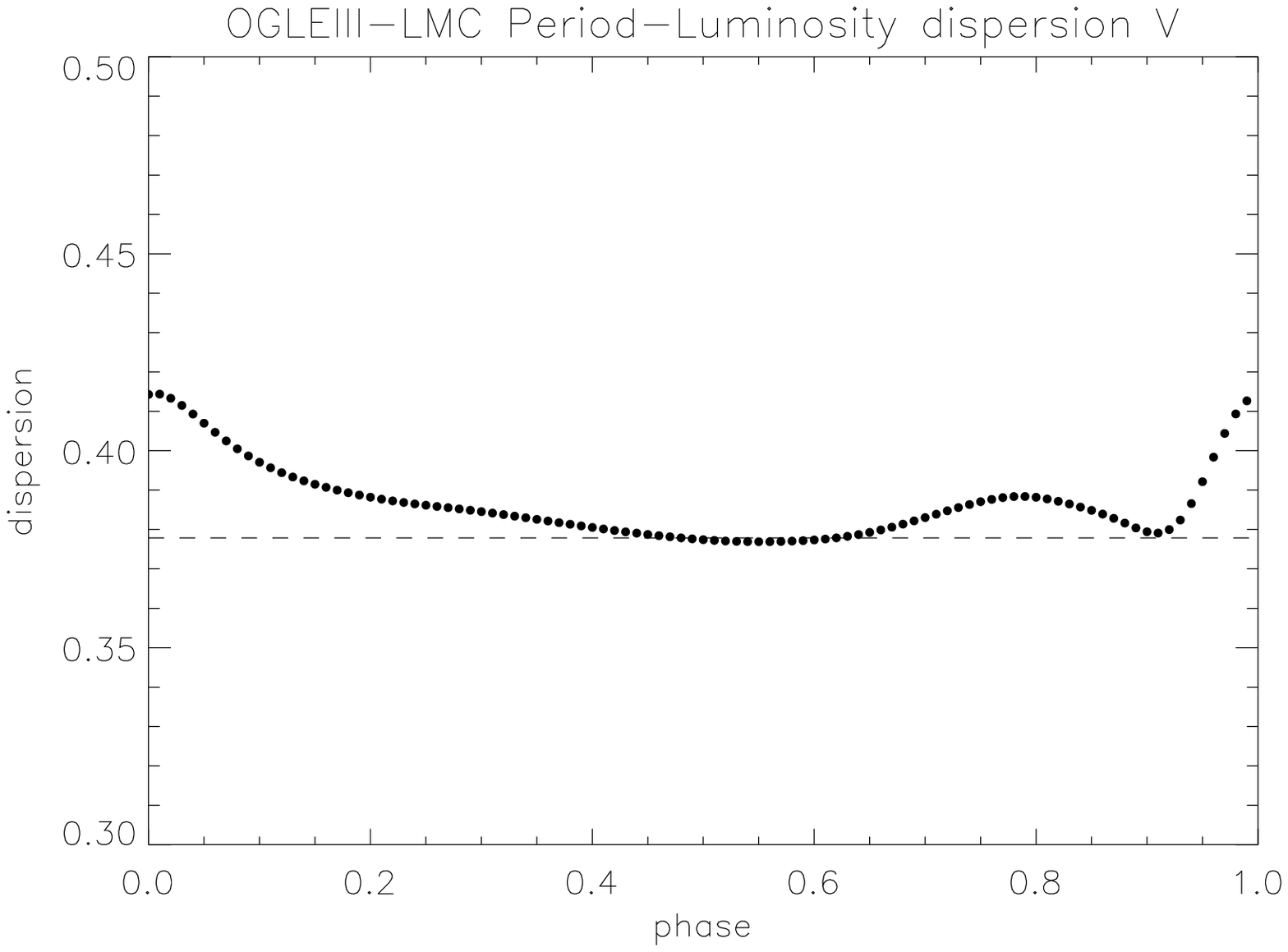}
\includegraphics[width=\columnwidth]{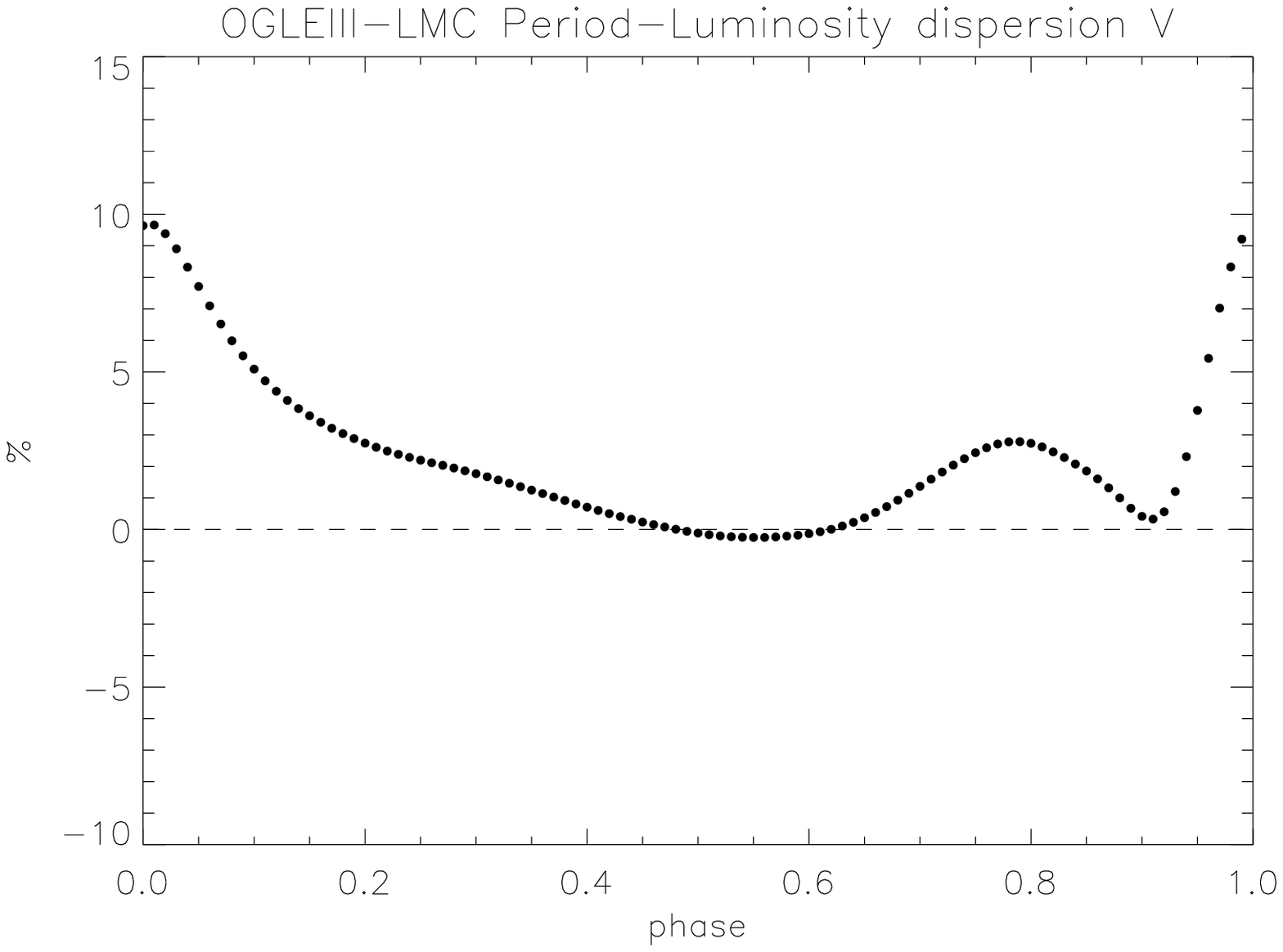} \\
\includegraphics[width=\columnwidth]{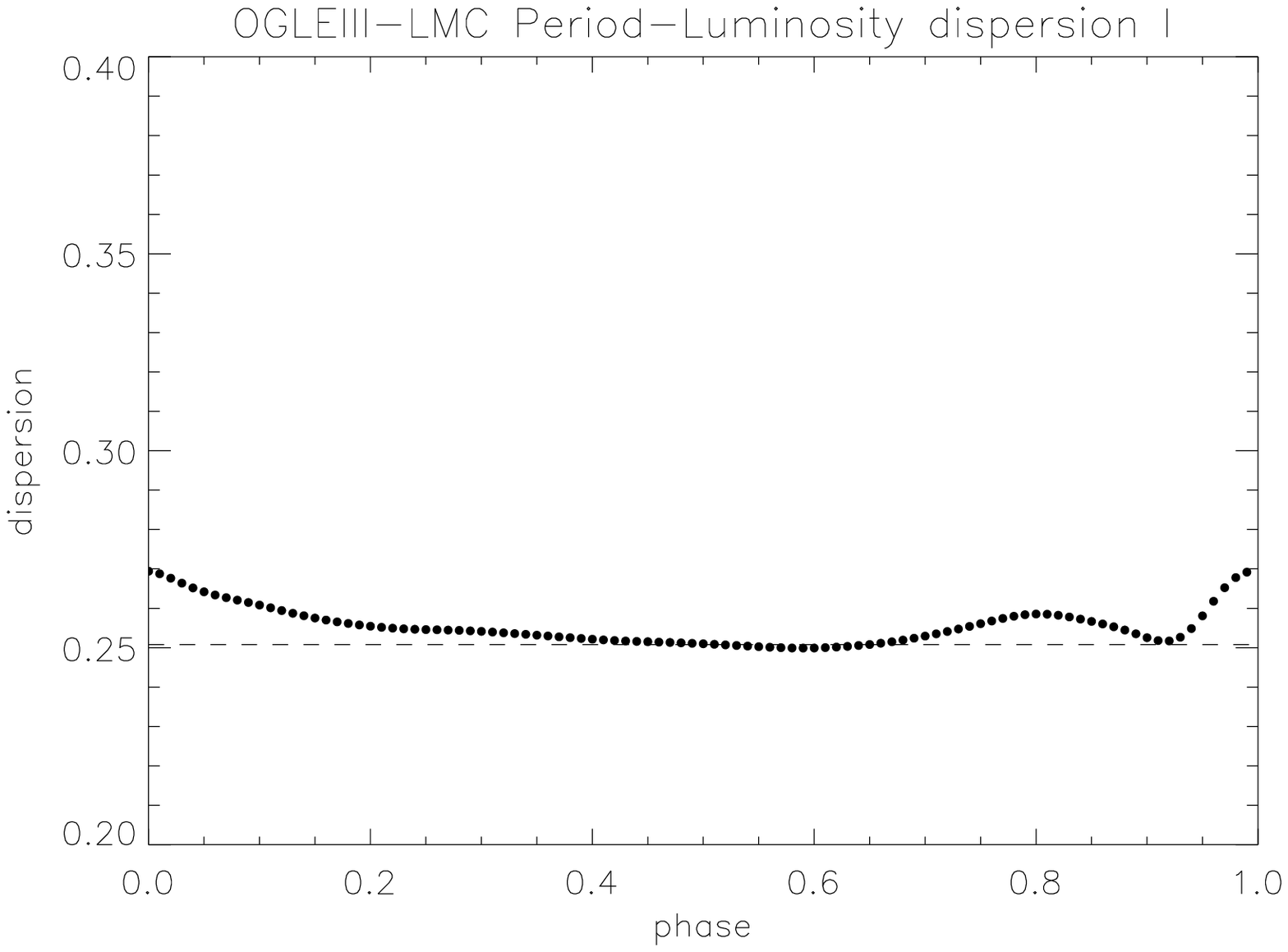}
\includegraphics[width=\columnwidth]{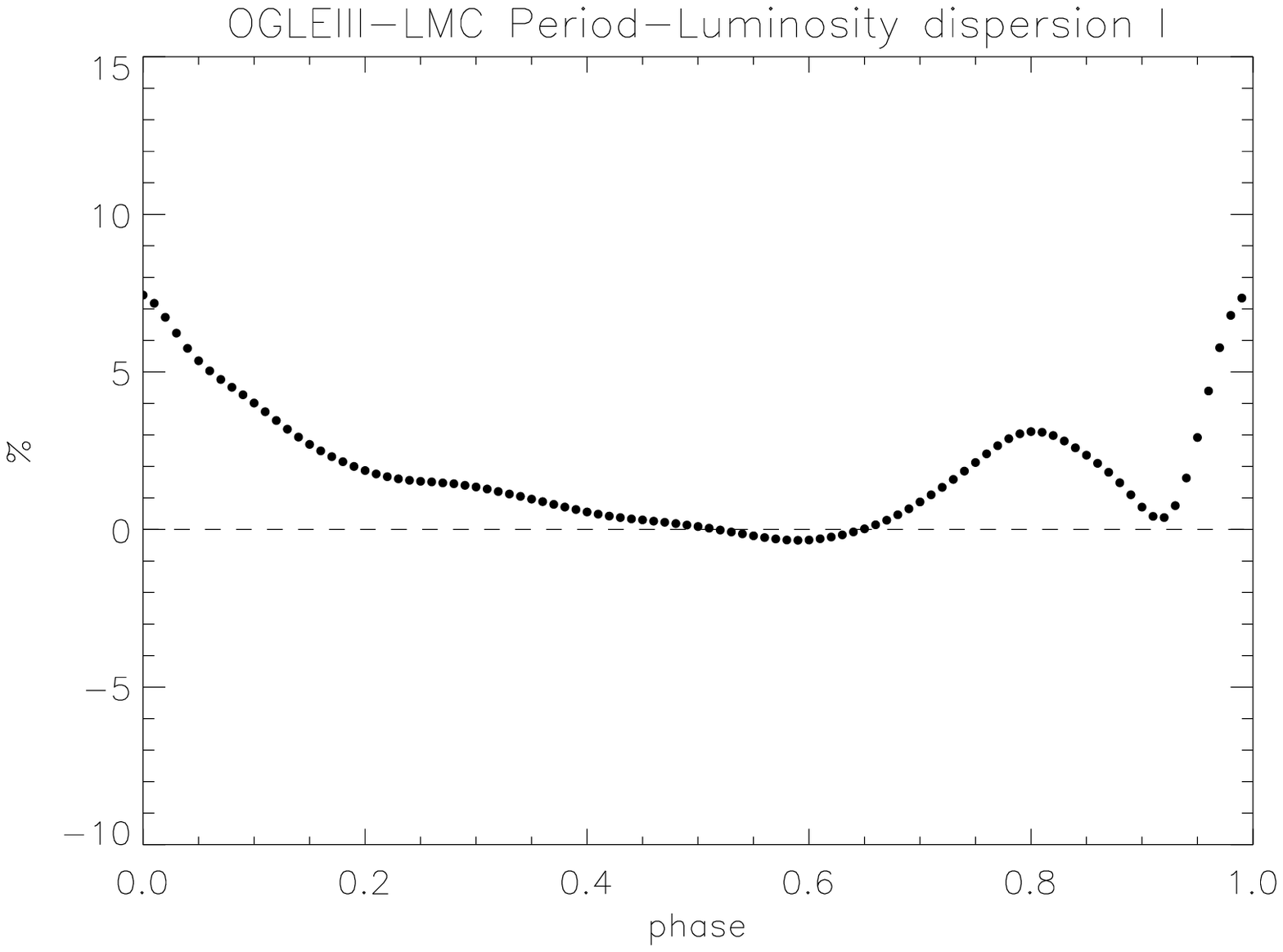} \\
\includegraphics[width=\columnwidth]{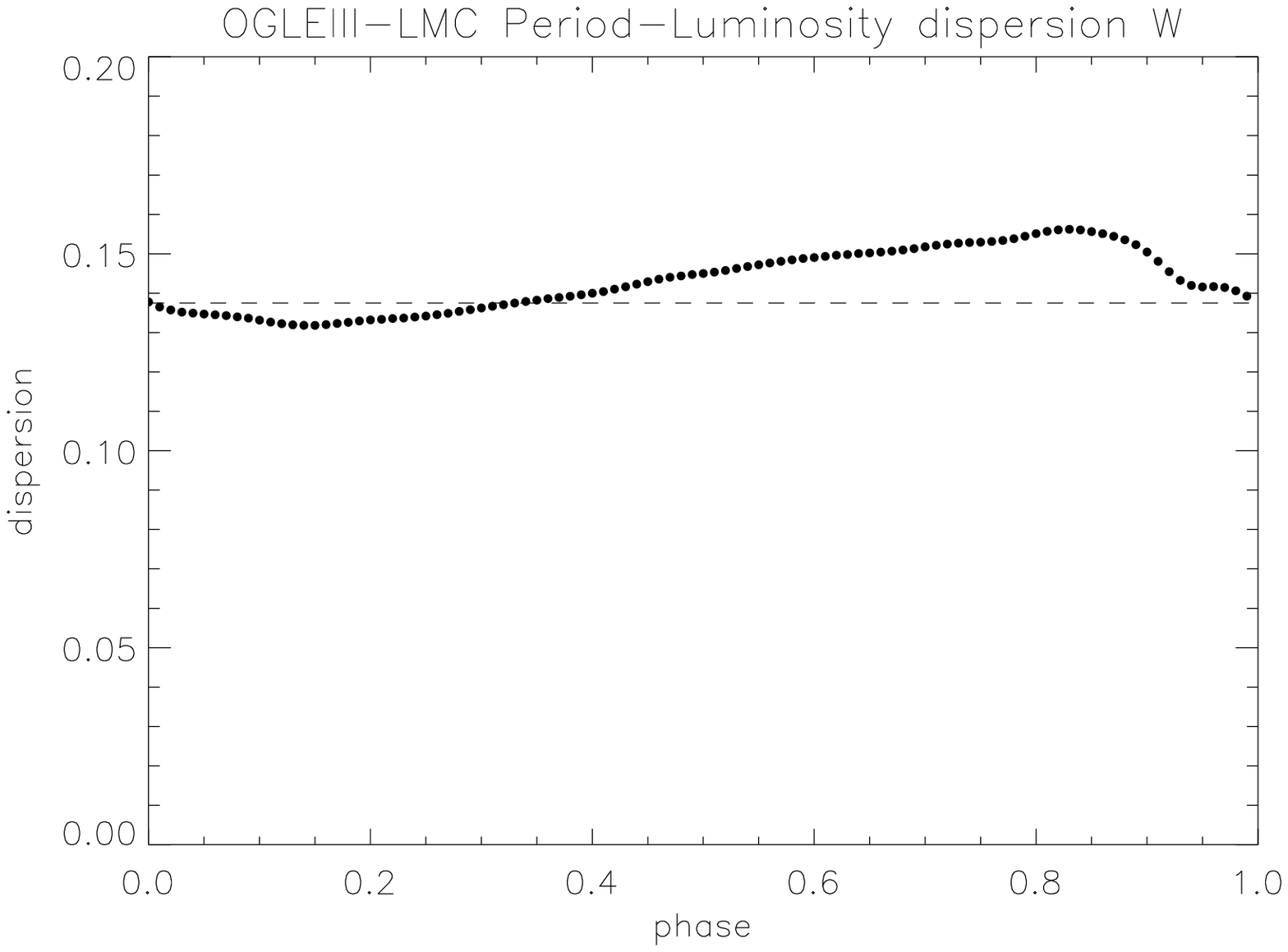}
\includegraphics[width=\columnwidth]{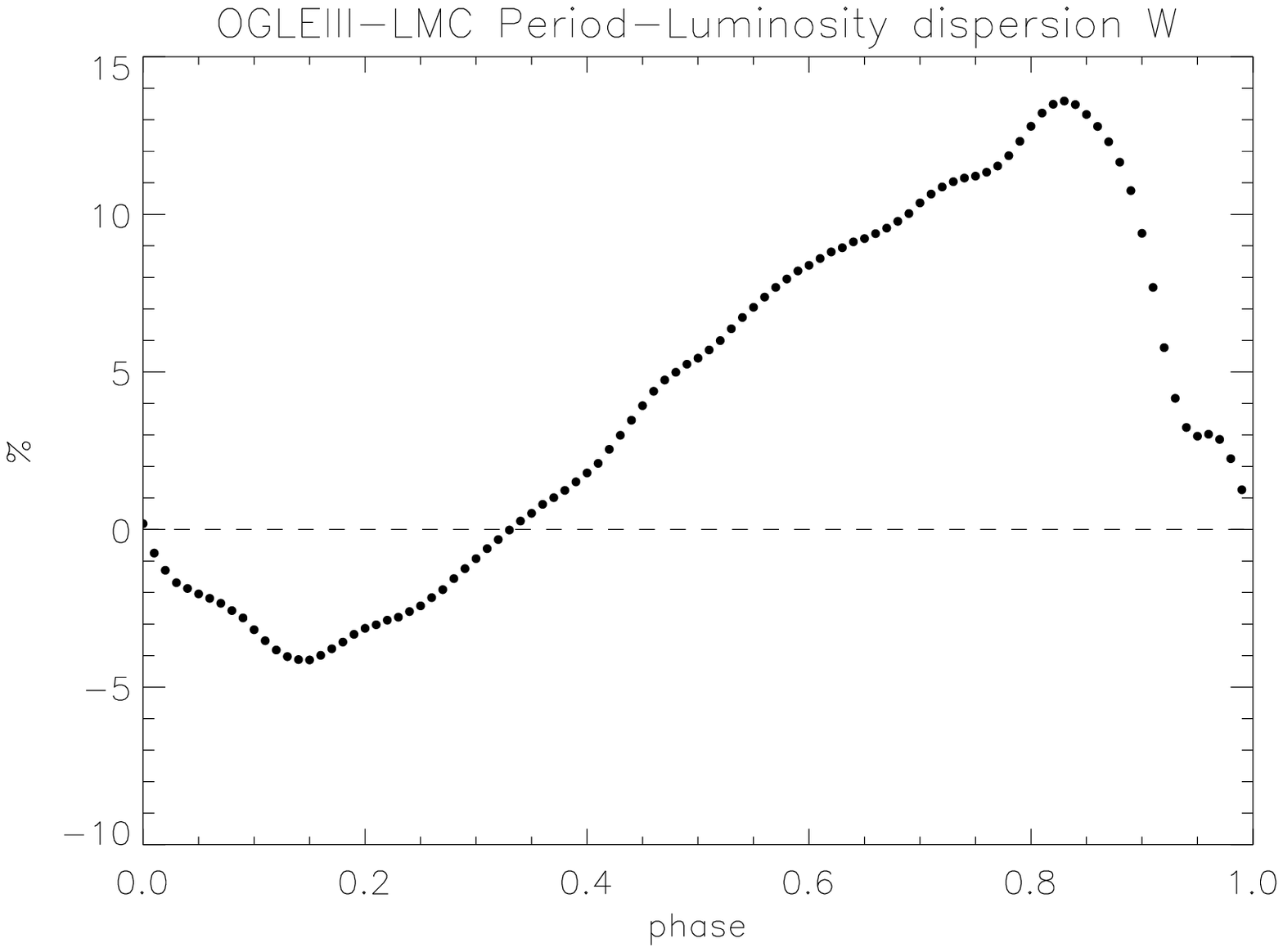}
\caption{{\bf (a) Left:} Dispersions of the multi-phase P-L relations as a function of pulsational phase. The dashed lines are the dispersions from the mean light P-L relations. {\bf (b) Right:} Percentage change of the dispersions when compared to the mean light P-L dispersions. Negative percentage means the dispersion is smaller than the mean light P-L dispersion.} 
\label{fig:7}
\end{figure*}

It is straightforward to calculate the dispersion of the multi-phase P-L relations as a function of pulsational phase. We use the LMC multi-phase P-L relations as an demonstration in this sub-section. Short and long period multi-phase P-L relations were used when calculating the overall dispersions. Results for $VIW$ band multi-phase P-L relations are shown in left panel of Figure \ref{fig:7}. In this Figure, dispersions from mean light P-L relations were included for comparison. Right panels of Figure \ref{fig:7} present the percentage change of the dispersions from the multi-phase P-L relations when compared to the mean light P-L dispersions.  

From Figure \ref{fig:7} it can be seen that minimum dispersion occurs at phase $\sim0.55$, $\sim0.59$ and $\sim0.14$ in the $VIW$ band, respectively. However, these dispersions are close to the dispersions from mean light P-L relations, with the largest difference being $\sim4$\% in the $W$ band. This result implies that the dispersion at mean light P-L relation is comparable to the minimum dispersion from multi-phase P-L relations. Hence, the applicability of mean light P-L relation in distance scale works is reinforced. Interestingly, the largest dispersion occurs at maximum light for both the $V$ and $I$ band.

\section{Discussion and Conclusion}\label{s:4}

In this paper, we present two new aspects in Cepheid P-L relation research: the study of mid-infrared and multi-phase P-L relations. Both of these studies utilized the LMC and SMC Cepheids catalogs from OGLE-III.

The MIR P-L relations were derived for Cepheids in both Magellanic Clouds using {\it Spitzer} archival data. These MIR P-L relations were also applied to derive the distance moduli to IC 1613 and NGC 6822, showing a good agreement with published distances. When comparing the empirical P-L slopes for LMC and SMC Cepheids, as listed in Table \ref{tab:1}, these P-L slopes suggested that they could be independent of metallicity, at least for metallicities bracketed by these two low-abundance galaxies. This is in contrast with the study of the synthetic P-L relations from \citet{nge2011}, which found that the synthetic MIR P-L slopes could be dependent on metallicity, suggesting a need for future work. Comparisons of the empirical and synthetic P-L slopes show that the LMC P-L slopes agree well with synthetic P-L slopes from the $Z=0.008$ model set. The empirical SMC P-L slopes also show a better agreement to the synthetic P-L slopes from the same model set as in LMC.

The multi-phase P-L relation for Cepheids in the SMC was investigated for the first time and compared to the LMC. These multi-phase P-L relations not only revealed that P-L relations, at least in the $VIW$ bands, are dynamic within the cycles of pulsations, but also behave differently for LMC and SMC Cepheids. This could be due to the metallicity difference of these two galaxies. Minimum dispersions occurs at specific phases for the $VIW$ band multi-phase P-L relations; however, these minimum dispersions do not differ significantly from the dispersions obtained from the mean light P-L relations. Of particular interest is the ``anomalous'' behavior of the multi-phase P-L relations in the phases between $\sim0.7$ and $\sim 0.9$, as evident from Figures \ref{fig:5} to \ref{fig:7}. This may be due to the interaction of the hydrogen ionization front (HIF) and the stellar photosphere, which are not always co-moving during a stellar pulsation cycle and can engage/disengage at various phases and/or period ranges, or the presence of shock in photosphere at these phases.

\acknowledgments
CCN thanks the funding from National Science Council (of Taiwan) under the contract NSC 98-2112-M-008-013-MY3.

\end{document}